\documentclass[conference]{IEEEtran}
\usepackage{amsmath,amsfonts}
\usepackage{algorithmic}
\usepackage{algorithm}
\usepackage{array}
\usepackage[caption=false,font=normalsize,labelfont=sf,textfont=sf]{subfig}
\usepackage{textcomp}
\usepackage{stfloats}
\usepackage{url}
\usepackage{verbatim}
\usepackage{graphicx}
\usepackage{cite}
\usepackage{fancyhdr}
\def\BibTeX{{\rm B\kern-.05em{\sc i\kern-.025em b}\kern-.08em
    T\kern-.1667em\lower.7ex\hbox{E}\kern-.125emX}}

\begin{document}
\raggedbottom

\title{zkDFL: An efficient and privacy-preserving decentralized federated learning with zero-knowledge proof}

\author{
\IEEEauthorblockN{1\textsuperscript{st} Mojtaba Ahmadi}
\IEEEauthorblockA{\textit{Shahid Beheshti University} \\
Tehran, Iran \\
mojtaba27ahmadi@gmail.com}
\and
\IEEEauthorblockN{2\textsuperscript{nd} Reza Nourmohammadi}
\IEEEauthorblockA{\textit{The University Of British Columbia} \\
Kelowna, Canada \\
reza.nourmohammadi@ubc.ca}
}

\maketitle

% \thispagestyle{fancy}
% \fancyhf{}
% \renewcommand{\headrulewidth}{0pt} % Remove header rule line

% \lhead{3rd IEEE International Conference on AI in Cybersecurity (ICAIC)}
% \lfoot{979-8-3503-8185-6/24/\$31.00 ©2024 IEEE}

\pagestyle{plain}

\begin{abstract}
Federated learning (FL) has been widely adopted in various fields of study and business. Traditional centralized FL systems suffer from serious issues. To address these concerns, decentralized federated learning (DFL) systems have been introduced in recent years. With the help of blockchains, they attempt to achieve more integrity and efficiency. However, privacy preservation remains an uncovered aspect of these systems. To tackle this, as well as to scale the blockchain-based computations, we propose a zero-knowledge proof (ZKP)-based aggregator (zkDFL). This allows clients to share their large-scale model parameters with a trusted centralized server without revealing their individual data to other clients. We utilize blockchain technology to manage the aggregation algorithm via smart contracts. The server performs a ZKP algorithm to prove to the clients that the aggregation is done according to the accepted algorithm. Additionally, the server can prove that all inputs from clients have been used. We evaluate our approach using a public dataset related to the wearable Internet of Things. As demonstrated by numerical evaluations, zkDFL introduces verifiability of the correctness of the aggregation process and enhances the privacy protection and scalability of DFL systems, while the gas cost has significantly declined.
\end{abstract}


\section{Introduction}
Machine learning has become a major focus of research and businesses in recent years \cite{mahesh2020machine}. Current machine learning implementations face challenges such as the cost of data gathering, training computations, and communication overhead \cite{baier2019challenges}. Another critical issue is safeguarding the privacy of private data when utilizing clients’ data \cite{liu2021machine}. Federated learning (FL) is a machine learning framework designed to tackle these issues \cite{mcmahan2017communication}. In FL systems, privacy is preserved by delegating the training process to clients. Specifically, each client trains its local model with its own data and sends the local model instead of the data itself. The server aggregates the models from all clients to obtain an updated global model. This approach not only addresses privacy concerns but also eliminates the need to transmit large amounts of data to a central server.\\
FL systems still include one central server that is honest-but-curious and can see all clients’ data. They can enhance privacy by adding noise to data, as reported in \cite{el2022differential}. On the other hand, although FL frameworks have significant features, the clients still need to assume that the server is honest and aggregates correctly, while the aggregation process is not done in a transparent way. This can lead to a global aggregation attack in which, for instance, the server might not use all clients' data to save computation cost or even weight some clients' models higher than others.\\
To address the concern of a central server, blockchain-based FL systems are introduced for both permissioned blockchains \cite{li2020blockchain,lu2020low}, and public blockchains \cite{weng2019deepchain,martinez2019record}. Generally, they aim to delegate the server’s tasks to a smart contract on-chain. While there is a variety of studies on merging blockchain with FL, not enough research has been conducted on handling large-scale models using smart contracts. For instance, CNN models can have over millions of parameters that cannot be stored in the blockchain. High gas cost of blockchain due to heavy computations \cite{eberhardt2018zokrates} in addition to the visibility of clients’ data, is a major obstacle in using the blockchain for real-world FL applications.\\
In this paper, we propose a novel federated learning framework for handling large-scale models, using a zero-knowledge proof system for proving the correctness of the aggregation algorithm and also for avoiding the leakage of clients’ model data. Furthermore, with the assistance of blockchain and smart contracts, clients can verify the correctness of the total process.\\
In summary, our contributions are as follows:
\begin{itemize}
\item{We propose a decentralized federated learning framework in which the central server aggregates clients’ local models. Then, using a ZKP algorithm, the server proves the correctness of the aggregation.}
\item{To hide the clients' models from others, the server calculates the hashes of the models and proves that the hash values have been calculated correctly. Also, verification of the proofs is performed by the deployed smart contracts, and the clients are able to check the correctness of the claimed computations.}
\item{This framework utilizes blockchain to ensure integrity and can deal with large-scale models via a verifiable central server as well.}
\end{itemize}
The remainder of the paper is structured as follows: The next section covers the preliminaries of our work. In Section 3, we provide a literature review. In Section 4, we delve into our proposed framework, and in the final section, we evaluate the proposed framework from different aspects, presenting the simulation results.

\section{Preliminaries}
\subsection{Blockchain and smart contract}
There are several types of blockchain implementations. In 2008, the first decentralized and transparent payment system, Bitcoin, was introduced \cite{nakamoto2008bitcoin}. Ethereum \cite{buterin2014next} the second-largest blockchain after Bitcoin, has the capability of deploying smart contracts. In fact, a smart contract \cite{bogner2016decentralised} is a program running on the blockchain. Each smart contract has its own unique address and its code is visible to everyone. The first implementation of smart contract was on Ethereum using Solidity language. Smart contracts are independent of third parties and are executed when the conditions written in their code are met \cite{zheng2020overview,zou2019smart}.

Nowadays, blockchains have been widely used in different fields of technology such as healthcare \cite{agbo2019blockchain,attaran2022blockchain}, the Internet of Things \cite{wang2020blockchain,dorri2017towards}, and energy \cite{andoni2019blockchain,bao2020survey}.
The main properties of blockchain are:
\begin{itemize}
    \item {\bf{Decentralization:}} blockchain uses the peer-to-peer network, so there is no need for a third party which means that the central nodes are removed. 
    \item {\bf{Immutability:}} data stored on  the blockchain are permanent and hardly can be altered.
    \item {\bf{Traceability:}} all data and transactions on blockchain can be traced back to the source.
\end{itemize}

Blockchains are divided into three major categories according to their nodes' interactions:
\begin{itemize}
\item{\bf{Public Blockchain:}} 
In public blockchains, anyone can join and make transactions or even participate in the consensus process. For example, Bitcoin and Ethereum are public blockchains. Transactions are processed by a number of nodes, and the speed is limited. These blockchains are more suitable for large-scale federated learning with heavy computations, but it can be more challenging to identify malicious nodes.
\item{\bf{Private blockchain:}}
Private blockchains, also known as permissioned blockchains, are controlled by a central authority that manages which nodes can participate in the network. Although all nodes’ activities are visible to each other, decentralization is not completely achieved. Private blockchains are more suitable for small-scale federated learning, and malicious node attacks are less likely.
\item{\bf{Consortium blockchain:}}
Consortium blockchains are controlled by multiple organizations or nodes, and these nodes are authorized to generate new blocks. They are a type of private blockchain in which different nodes have authority instead of one central authority.
\end{itemize}

\subsection{Zero-Knowledge Proof}

 Zero-knowledge proof (ZKP) is a protocol between a prover and  a verifier, in which the prover attempts to convince the verifier about the validity of a statement. Each ZKP protocol needs to satisfy three conditions:
 \begin{itemize}
     \item {\bf{Completeness:}} If the prover and verifier follow the protocol correctly, the honest verifier accepts the validity of a true statement.
     \item {\bf{Soundness:}} The probability that a cheating prover can convince the verifier of a false statement is negligible.
     \item {\bf{Zero knowledge:}} The verifier learns nothing from the protocol except the validity of the statement.
 \end{itemize}

There are several parameters that indicate the efficiency of a zero-knowledge proof (ZKP) algorithm, including proving time, verification time, proof size, and interaction rounds. The earliest ZKPs only achieved zero knowledge in the presence of a computationally bounded verifier, but in \cite{brassard1988minimum} Brassard et al. showed that proofs against an unbounded verifier can also hold zero knowledge. 
%One of the most important researches \cite{gmr88}, is the first sound, complete, and zero-knowledge interactive proof systems.
Sumcheck protocol, on the other hand, is another type of interactive protocols. In \cite{lund1992algebraic}, a sumcheck protocol for delegating computation from verifier to prover was proposed. In \cite{goldwasser2015delegating}, an interactive proof protocol is proposed for layered arithmetic circuits, known as the GKR protocol. 
Generally, GKR has been the subject of many research projects such as \cite{xie2019libra,zhang2020transparent,liu2021zkcnn}.

On the other hand, the first non-interactive zero-knowledge (NIZK) proofs were developed by \cite{blum1991noninteractive}. The non-interactive characteristic was important in many practical situations as there is no need to interact  with the prover to verify the proof every time. 
In \cite{groth2010short}, the first constant-size NIZK arguments system in introduced. By reducing an NP-statement to Quadratic Span Program (QSP) and Quadratic Arithmetic Program (QAP), Gennaro introduced a new design in which proof size and complexity were so smaller than previous NIZK systems \cite{gennaro2013quadratic}. Additionally, the QAP idea was employed in the first practical protocol, the Pinocchio protocol, by Parno et al. \cite{parno2016pinocchio}. Improvements to the Pinocchio protocol led to a new protocol, called Groth16, with a smaller proof size.

In the field of non-interactive proofs, a more interesting concept is SNARK (Succinct Non-interactive Argument of Knowledge). If SNARK systems reveal no information about the intermediate steps (the witness), they are known as zkSNARK. These algorithms have been widely used, especially in blockchain, for scaling and securing transactions \cite{hopwood2016zcash,eberhardt2018zokrates,polygon,aztec,zksync}.

\subsection{Groth16}
Groth16 is a zkSNARK protocol based on a set of polynomial equations and employs symmetric pairing cryptography. In this protocol, the prover computes a number of grouped elements, and the verifier checks pairing product equations. Moreover, in this protocol, any program is converted into a quadratic arithmetic circuit. The corresponding algorithm consists of three phases:\\
\begin{enumerate}
    \item Generating common reference string (CRS), containing proving key $p_{k}$ for the prover and verification key $v_{k}$ for the verifier based on a given relation. It is run by a trusted third party. In addition, the common parameters of the proof system, such as selecting elliptic curve group and defining the bilinear pairing function, are generated in this step.
    \begin{equation}
        CRS \gets Setup(R):
    \end{equation}    
    \item  Generating a proof $\pi$, which reveals no information, based on public input and witness data using a series of mathematical computations.
    \begin{equation}
        \pi\gets Prove(p_{k},x,w):
    \end{equation}    
    \item Verification of the proof, $\pi$, by the verifier. It takes the public input, the CRS, and the proof and returns a boolean output.  
    \begin{equation}
        0,1\gets Verify(v_{k},x,\pi):
    \end{equation}    
\end{enumerate}

\subsection{Trusted setup}
zkSNARK protocols, like Groth16, rely on a Common Reference String (CRS) as a public parameter for proving and verifying. This CRS must be generated by a trusted third party before executing the proving or verifying algorithms to prevent potential attacks. Groth16 is not universal, and proofs are specific to a given program. By changing the program, the CRS needs to be regenerated. The CRS is generated in two phases. The first phase is known as the 'Power of Tau' ceremony. The output can be used for all circuits. The second phase takes the output of the previous phase and converts it into a circuit-specific CRS. \cite{bowe2017scalable}.

Setup ceremonies are based on multi-party computation (MPC). Typically, there are multiple parties that participate in the ceremony randomly, and each party adds some randomness to the output of the previous party. All parties must keep their inputs private and delete the computed amounts immediately after the ceremony. The key factor of this ceremony is that if only one party is honest, the output is reliable.
\subsection{Hash function}
There are a range of hash functions like SHA family \cite{burrows1995secure} and Keccak \cite{bertoni2009keccak}. Although the Keccak hash function is very secure and can be implemented in hardware or software efficiently, it is not suitable for SNARK-based algorithms. In fact, these hash functions contain a variety of operators that are hard to convert to SNARK circuits. To address the need for hash functions in zero-knowledge proof algorithms, some new hash functions have been developed. One of the simplest and most efficient SNARK-friendly hash functions is MiMC \cite{albrecht2016mimc}.

\begin{figure*}[!t]
\centering
\includegraphics[width=5in]{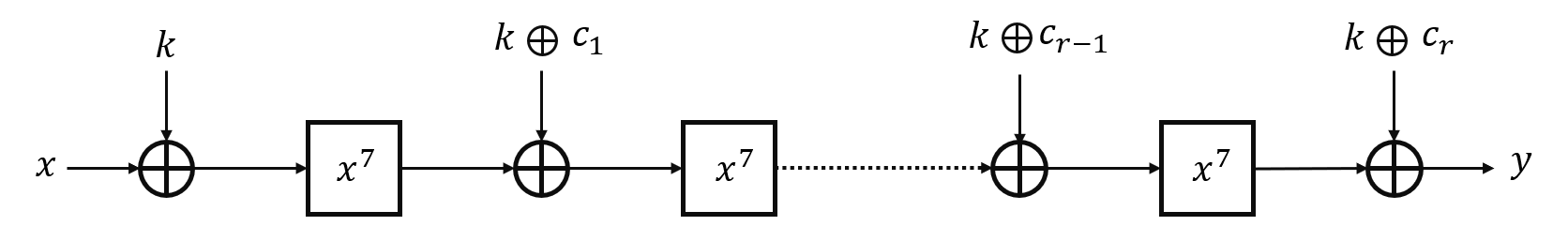}
\caption{MiMC7 Structure.}
\label{mimc}
\end{figure*}

In our work we use MiMC7 hash function which contains r rounds. The core part of MiMC7 is $f(x)=x^{7}$. Let $\mathbb{F}_{q}$ be the finite field with $q$ elements. So, the function is computed in $\mathbb{F}_{q}$. A uniformly random secret key $k\in \mathbb{F}_{q}$ and a round constant $c_{i}\in \mathbb{F}_{q}$ is added to plaintext in each round. The encryption process is defined as
\begin{equation}
E_{k}(x)=(F_{r-1}\circ F_{r-2} \circ ... \circ F_{0})(x)+k,
\end{equation}
where $x\in \mathbb{F}_{q}$ is the plaintext, r is the number of rounds, $F_{i}$ is the round function for round $i\ge 0$, and $k\in \mathbb{F}_{q}$ is the key. Each $F_{i}$ is defined as
\begin{equation}
F_{i}(x)=(x+k+c_{i})^7,
\end{equation}
where $c_{i}\in \mathbb{F}_{q}$ are the round constants and $c_{0}=0$.\\
The random constants $c_{i}$ do not need to be generated every time and are chosen at the instantiation of MiMC7. The secret key k is the same in each round. The MiMC7 structure is illustrated in Figure \ref{mimc}.

\section{literature review}
The concept of federated learning was first introduced by Google \cite{mcmahan2017communication} in 2016. In that framework, users update the model parameters trained locally on their mobile phones and upload the parameters to the cloud, training the centralized model alongside other users. The main idea is to aggregate local models learned on multiple clients without sending the private information. After that, a body of research has focused on improving integrity, privacy, incentive mechanisms, and performance. 

A range of implementations for supporting effective machine learning models has been introduced recently. For instance in \cite{yurochkin2019bayesian} and \cite{ryffel2018generic},  deep neural networks and in \cite{chen2018privacy} logistic regression models are implemented in federated frameworks, while
in \cite{lin2017deep} Deep Gradient Compression is utilized to reduce the communication bandwidth in large-scale systems.

In \cite{li2020privacy}, \cite{sun2020federated}, and \cite{aono2017privacy} the privacy-preserving issue is addressed by multi-party computation, differential privacy (DP) and homomorphic encryption respectively. 

In \cite{yang2023fedzkp} a federated learning framework called FedZKP is proposed, with the ability of secure ownership verification using ZKP. FedZKP algorithm is guaranteed to defeat a variety of existing attacks without disclosing private data.

In \cite{nguyen2023preserving} Nguyen and his coworkers developed a framework which is empowered by homomorphic and ZKP module for secure aggregation and defence mechanism which acts against poisoning attacks. These attacks can occur when the central server is an adversary trying to infer the membership information of clients by inspecting their local models’ weights.

In \cite{zhao2021veriml}, Zhao et al have introduced a framework called VeriMl for outsourcing machine learning. VeriMl validates neural network predictions by using SNARK. The prover commits to the values of all intermediate layers, and the verifier validates one layer randomly. Although it verifies the integrity of machine learning computation and the correctness of predictions, it dose not guarantee the user data and model confidentiality. It also justifies the challenge of the overhead on the prover time, as it is too expensive to apply SNARK to the entire machine learning model.

Following the emergence of blockchain, several studies attempted to integrate blockchain with federated learning. The goal was to address some of the issues raised in centralized federated learning.

A blockchain-based trust management system, proposed in \cite{kang2019incentive}, was designed for a reliable federated learning (FL) algorithm in which the server selects reliable and high quality data owners as FL clients by motivating participation of reliable end-devices. In fact, this paper defines a reputation-based metric to determine the reliability of end-devices through a multi-weight subjective logic model. The authors proposed an incentive mechanism based on contract theory to motivate the participation of end-devices in federated learning. This reputation relies on the clients' previous interactions and behaviors. They also utilize blockchain to achieve secure reputation management, motivating the clients.

BAFFLE \cite{ramanan2020baffle} is a decentralized framework without the need for a central aggregator, capable of achieving the same training results as conventional federated learning systems. Baffle uses blockchain to store and share the global model, employing aggregation with smart contracts. Blockchain is utilized as an immutable ledger to record local gradients and aggregate them in a trusted manner, allowing local devices to update the global model parameters for their respective local models. However, their approach has several limitations related to gas costs and the data size of each blockchain-based transaction, and these costs are all transferred to the nodes. When all nodes have to perform consensus tasks, the computational overhead is substantial.

In \cite{ruckel2022fairness} a fair blockchain-based federated learning framework is introduced. This framework can validate the correctness of the clients' local model training, enabling verification of the training process's correctness. In fact, it attempts to simultaneously address the challenges of achieving fairness, integrity, and privacy preservation for all clients by utilizing blockchain, local differential, and zero-knowledge proofs. LDP with Laplacian noise ensures clients’ model weights cannot leak within their private data, and non-interactive ZKPs are used to prove the integrity of the training and evaluation process, which is required for providing fair incentives managed by a smart contract. It incentivizes each client based on their individual contribution to the global model. On the other hand, their system only supports linear regression, which is not a practical achievement for federated learning systems.

\cite{heiss2022advancing} proposed a framework in which each client can prove that the learning process has been done according to the committed algorithm by utilizing both ZKP and blockchain. In this paper, the training process is converted to ZKP circuits, and the verification of the correctness of training is conducted by a smart contract. As the computation cost of this algorithm is very high, a single-layer perceptron model is simulated in this paper, and the dataset is converted to a simpler version to lessen the amount of computations. With this method, large-scale models cannot be implemented due to high computations, long proof generation time, and large smart contract size.

Based on existing research, we propose a system that addresses the problems of both centralized and decentralized FL systems. In centralized FL systems, there is no way to be sure that the server performs all computations correctly. In decentralized FL systems, the privacy of clients' models is not preserved properly, and also the high gas cost issue required for large-scale models makes it impractical in many cases. Our proposed system proves to each client that the aggregation algorithm has performed correctly using ZKP. Furthermore, all clients can verify the proof in a decentralized manner with low gas cost.

\section{Proposed Scheme}
\subsection{Setup initialization}
The proposed system consists of one trusted central server, with K clients. We assume that all clients are honest and contribute to the learning process honestly without trying to poison the system. Each client has a public address on the blockchain, which is shared with other clients before starting the process. The global model's parameters, at the beginning point, are known to each client, and all of them try to train the known model. Each client has local data and is capable of conducting the training process on the model with its data. The central server is assumed to have enough power to compute all needed tasks correctly.

\subsection{Local training}
Each client attempts to train its model based on local data. The process of local training and obtaining updated weights based on gradients is called one epoch. Clients compute gradients and obtain local weights. The server selects m clients from all K clients to participate in aggregating local weights to obtain global weights. This process, known as a global round, allows each client to be trained over multiple epochs before contributing to global aggregation. By doing so, communication costs are reduced as clients perform computations locally, iterating local updates several times before submitting to the server, achieving a balance between computational and communication costs.

\subsection{Aggregation algorithm}
There are a variety of FL aggregation algorithms, most of which have been compared in \cite{nilsson2018performance,bonawitz2019towards}. In our proposed system, we use averaging aggregation, known as FedAvg, first introduced by Google in their FL aggregation strategy \cite{mcmahan2017communication}. FedAvg runs Stochastic Gradient Descent (SGD) in parallel on a fraction of clients, averaging the output of selected clients. When all clients’ inputs are submitted to the server, the server starts to aggregate the inputs. The pseudo-code of the aggregation algorithm is shown in Algorithm \ref{alg1}.

\begin{algorithm}[H]
\caption{FedAvg}\label{alg:alg1}
\begin{algorithmic}
\STATE
\STATE {\textsc{\textbf{server:}}}
\STATE \hspace{0.5cm} initiate $w_{0}$
\STATE \hspace{0.5cm} for each round do:
\STATE \hspace{1cm} $m\gets max(C.K,1)$
\STATE \hspace{1cm} select M clients randomly
\STATE \hspace{1cm} for each client $k\in M$ do:
\STATE \hspace{1.5cm} $w^{k}_{t+1}\gets ClientUpdate(w_{t})$
\STATE \hspace{1cm} $N\gets \sum_{k\in M}^{}n_{k}$
\STATE \hspace{1cm} $w_{t+1}\gets \sum_{k\in M}^{}\frac{n_{k}}{N}w^{k}_{t+1}$

\STATE {\textsc{$\bf{ClientUpdate}(w):$}}
\STATE \hspace{0.5cm} $\beta\gets$ local data divided to batches
\STATE \hspace{0.5cm} $w_{init}\gets w$
\STATE \hspace{0.5cm} for epoch e between 1 to E do:
\STATE \hspace{1cm} for batch $b\in \beta$ do:
\STATE \hspace{1.5cm} $w\gets w-\eta\nabla l(w;b)$
\STATE \hspace{0.5cm} return $w$ to server

\end{algorithmic}
\label{alg1}
\end{algorithm}

By changing the parameters C, the fraction of engaging clients; E, the number of local iterations; and B, the local minibatch size, we can control the amounts of computations.

\subsection{Proving the correctness of aggregation}
The server has to calculate the average amount of input weights and prove to clients that the total process has been done according to the claimed algorithm. The server initiates a zero-knowledge proof algorithm to prove to clients that the claimed FedAvg algorithm has been executed properly, ensuring the correctness of aggregation. The server utilizes the Groth16 algorithm to map the averaging algorithm to a SNARK circuit, wherein the computations are converted to Quadratic Arithmetic Program (QAP) statements. The trusted setup produces the proving key and verification key.

In this scenario, the server acts as the prover, and clients are verifiers. The prover runs the proving algorithm with the proving key and input parameters, finally producing the proof. Now it’s time to send the proof as well as public parameters to the verifiers. According to the Groth16 algorithm, some data are public. In our scenario, the outcome of the aggregation algorithm is considered as public data. On the other hand, we do not want to prove that we know some input data, the average amount of which is equal to the outcome. As a matter of fact, we need to prove that we use the same inputs and average them. To achieve this, we must consider this input data as public as well.

\subsection{Deploying smart contracts}
In order to verify the correctness of computations provided by the prover, two smart contracts, ($Contract_{h}$) for checking the summation of total hashes, and ($Contract_{p}$) for verifying the proof, will be deployed on the blockchain. Each client is able to call the contracts with its own input hash, verification key, proof, and public data to verify the summation and proof.

\subsection{Proving the utilizing of total inputs}
We have addressed the issue of proving that all input data was used. However, this scenario presents three major problems:

Firstly, if we need to send the input of each client to other clients, then there is no need for the aggregation algorithm to be performed by the server, and the clients are able to average these input data themselves, rendering the presence of the server meaningless.

Secondly, verifiers need to send all public data to the smart contract to prove the correctness of computations. For a considerable number of clients, the amount of data sent to the smart contract by clients would be huge, resulting in heavy communication costs. On the other hand, the smart contract may fail to support big input data, and the gas cost will rise significantly.

Thirdly, if clients send their input data to the blockchain, it will be visible to everyone. This can be a privacy concern and might make the system more vulnerable to attacks.

Considering all these concerns, we use the hash function to hide the input data of clients, diminish the size of smart contract inputs, and reduce the gas cost of execution on the blockchain.

In this situation, the server needs to calculate the hash of the input data and the average of them as well. Furthermore, the server must prove that it has calculated the hash of the same input data that was used in the aggregation algorithm. So, the server performs these two processes simultaneously and proves that it computes the average of the data and hashes the same data.

\subsection{Verifying the correctness of whole process}
The key point is that the server only sends the hash of the inputs and output as public data. In other words, when a verifier with these public data and proof attempts to verify the correctness of the computations, it can confirm that some input data has been hashed and averaged. In the end, to ensure that the mentioned hashes are exactly the hashes of the clients' input data, each client must calculate the hash of its input and compare it to the hash list of input data. By doing this, each client can verify that its input has been utilized and averaged.

To verify that all clients’ input data have been used, we deploy a smart contract in which each client sends the hash of its input data, and the smart contract adds these hashes together. It is clear that the server needs to sum the hash of input data as well and send the summation amount as public data. After the last client calls the smart contract, if the sum of the hashes calculated by the contract is equal to the summation amount claimed by the server, the entire process will be verified correctly, and the aggregation process will be completed.

\subsection{Final architecture}
\begin{figure}[!t]
\centering
\includegraphics[width=3.5in]{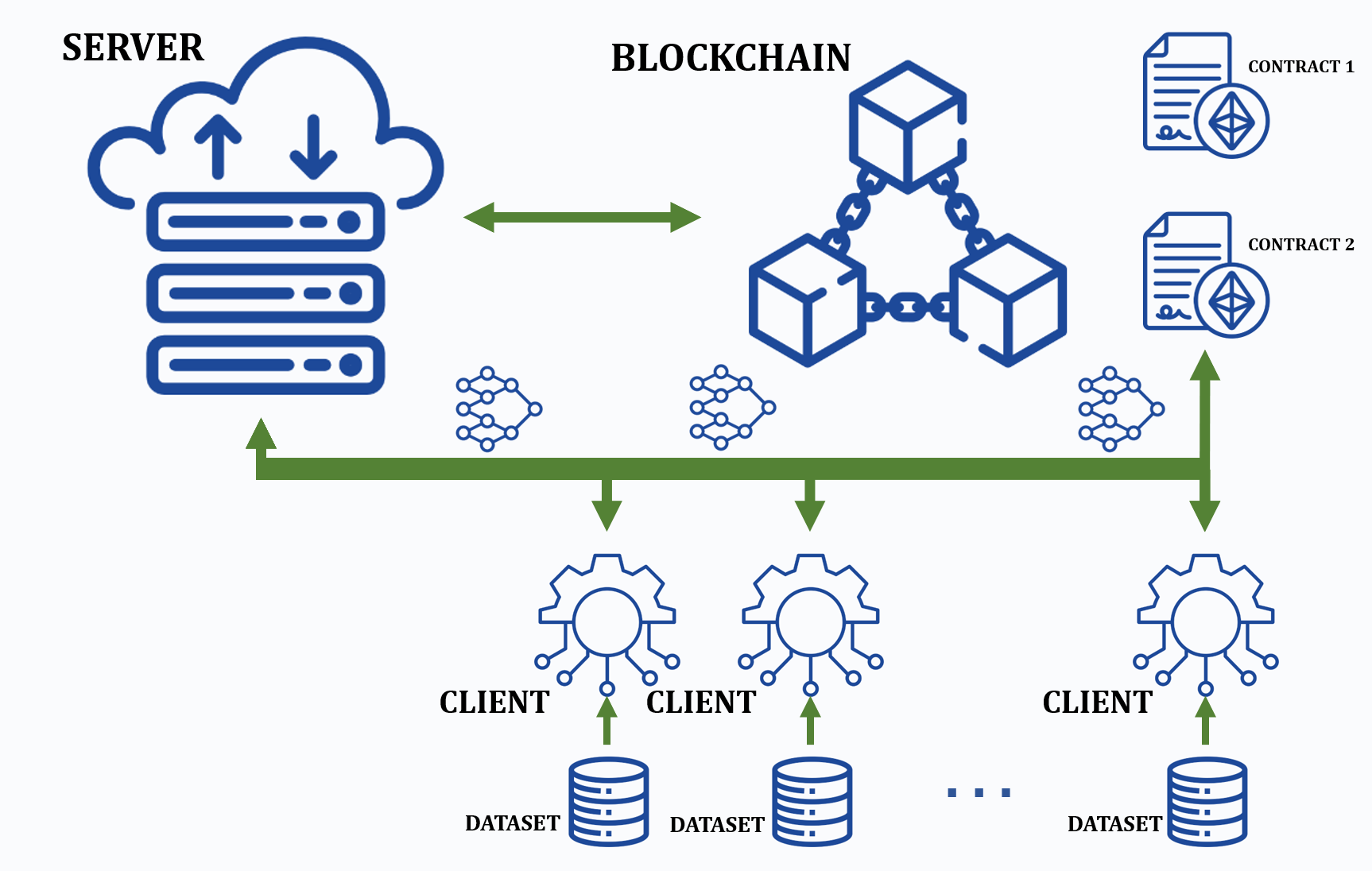}
\caption{Architecture of the proposed system.}
\label{framework}
\end{figure}

The architecture of the proposed system is shown in Figure \ref{framework}. In this system, the server chooses m clients randomly from the total clients. Each selected client trains its local model with local data using a predefined training algorithm in E epochs. The updated local models from the selected clients are sent to the server. Each client merges all weights and biases of all model layers into one matrix and then sends it to the server.

The server runs a ZKP algorithm in which the FedAvg algorithm operates on the clients' data and calculates hashes of all input data. It generates a SNARK circuit for these computations, and with the help of a trusted setup, the CRS needed for proving and verifying are generated.

After that, the server proves the correctness of hash calculations and averaging input data. When the proving process is finished, the server deploys one smart contract that allows the clients to verify the validity of hashes and another smart contract that verifies the proof of total computations.

Eventually, each client hashes its input data and sends the result to the first smart contract. In the contract, all input hashes are added together, and once the last client calls the smart contract with its own input hash, if the summation amount calculated by the smart contract is equal to what the server has claimed, the validity of the hashes is verified. After that, by calling the second smart contract, the proof is verified. The clients replace the global model with their local models, and the whole process continues until the stop criterion is met. The pseudo-code of the total process is indicated in Algorithm 2.

\section{EXPERIMENTAL ANALYSIS}
This section illustrates the performance analysis of the proposed zkDFL with a focus on the transaction cost and ZKP computation cost in decentralized federated learning. The experiments are executed five times, and the average of the measurements in each run is represented as the results.

\subsection{Experimental Setup}
\begin{figure}[!t]
\centering
\includegraphics[width=3in]{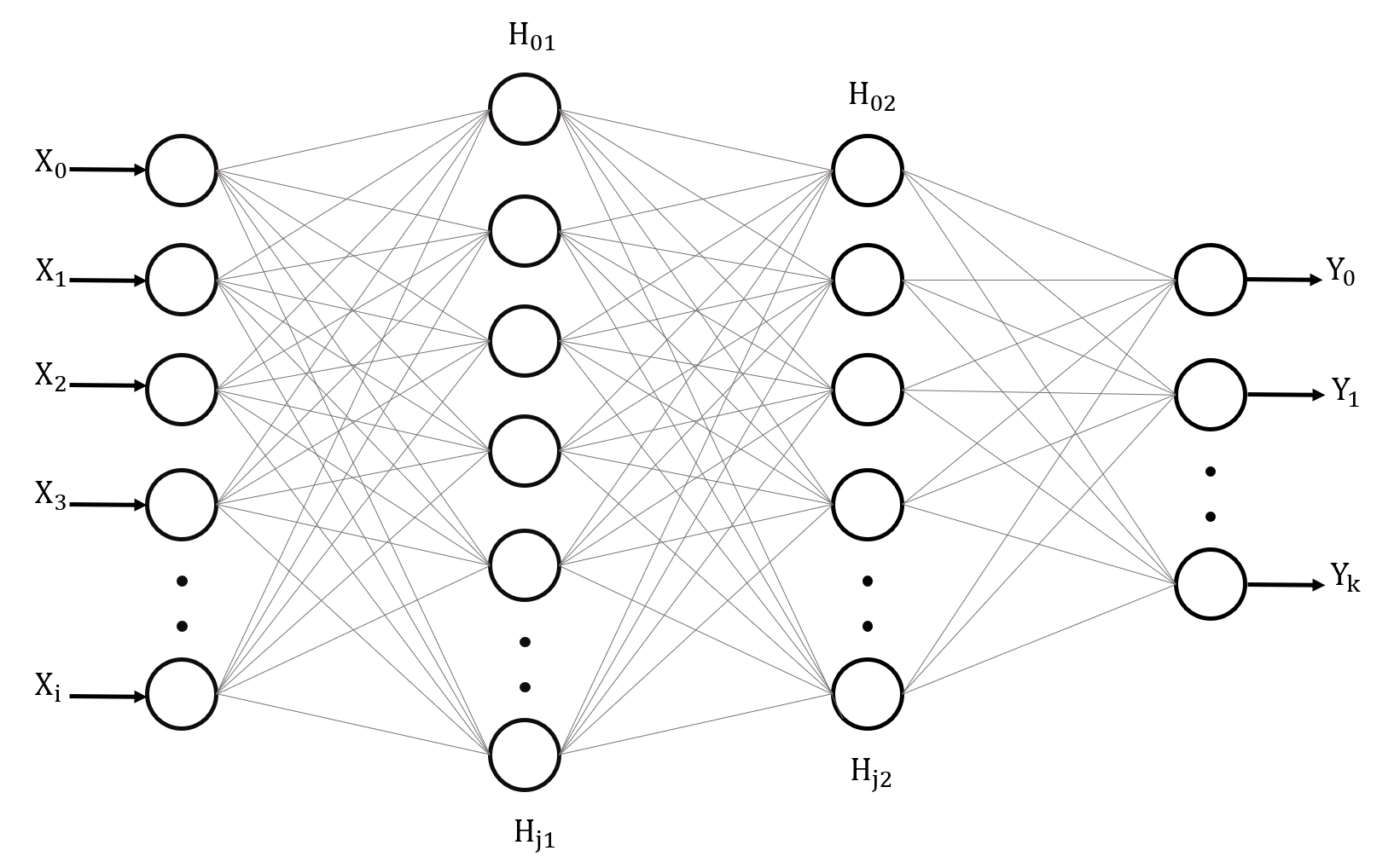}
\caption{MLP Architecture.}
\label{mlp}
\end{figure}

Model: we implement multi layer perceptron (MLP) neural network with different layers and neurons to investigate the impact of parameters size on performance. Figure \ref{mlp} shows a MLP model with two hidden layers. All the implementations are done in python3 using Pytorch library \cite{pyTorch}.

\begin{algorithm}[H]
\caption{zkDFL}\label{alg:alg1}
\begin{algorithmic}
\STATE
\STATE {\textsc{\textbf{server:}}}
\STATE \hspace{0.5cm} Initiate $w_{0}$
\STATE \hspace{0.5cm} for each round do:
\STATE \hspace{1cm} $m\gets max(C.K,1)$
\STATE \hspace{1cm} select M clients randomly
\STATE \hspace{1cm} $H_{sum}\gets 0$
\STATE \hspace{1cm} for each client $k\in M$ do:
\STATE \hspace{1.5cm} $w^{k}_{t+1}\gets ClientUpdate(w_{t})$
\STATE \hspace{1.5cm} $H^{k}\gets MiMC7(w_{t+1}^{k})$
\STATE \hspace{1.5cm} $circuit_{h}^{k}\gets SNARK(H^{k})$
\STATE \hspace{1.5cm} $H_{sum}\gets H_{sum}+H^{k}$
\STATE \hspace{1cm} $N\gets \sum_{k\in M}^{}n_{k}$
\STATE \hspace{1cm} $w_{t+1}\gets \sum_{k\in M}^{}\frac{n_{k}}{N}w^{k}_{t+1}$
\STATE \hspace{1cm} $circuit_{w}\gets SNARK(w_{t+1})$
\STATE \hspace{1cm} $w_{hash}\gets MiMC7(w_{t+1})$
\STATE \hspace{1cm} $circuit_{h}\gets SNARK(w_{hash})$
\STATE \hspace{1cm} $\pi\gets ZKP(circuit_{h}^{k},circuit_{w},circuit_{h})$
\STATE \hspace{1cm} $Contract_{h}\gets adding hashes )$
\STATE \hspace{1cm} $Contract_{p}\gets verify proof )$

\STATE {\textsc{$\bf{ClientUpdate}(w):$}}
\STATE \hspace{0.5cm} $\beta\gets$ local data divided to batches
\STATE \hspace{0.5cm} $w_{init}\gets w$
\STATE \hspace{0.5cm} for epoch e between 1 to E do:
\STATE \hspace{1cm} for batch $b\in \beta$ do:
\STATE \hspace{1.5cm} $w\gets w-\eta\nabla l(w;b)$
\STATE \hspace{0.5cm} return $w$ to server

\STATE {\textsc{$\bf{ZKP}(c1,c2,c3):$}}
\STATE \hspace{0.5cm} $CRS\gets$ trusted setup(c1,c2,c3)
\STATE \hspace{0.5cm} $\pi\gets Prove(CRS)$
\STATE \hspace{0.5cm} return $\pi$

\end{algorithmic}
\label{alg2}
\end{algorithm}

\subsubsection{Dataset}
 The dataset comprises motion sensor data of 19 daily and sports activities each performed by 8 subjects in their own style for 5 minutes \cite{misc_daily_and_sports_activities_256}. Sensors (x, y, z accelerometers, x, y, z gyroscopes, x, y, z magnetometers) were placed on each of five body parts (torso, right arm, left arm, right leg, left leg). Sensor units are calibrated to acquire data at 25 Hz sampling frequency. The number of features is 45 and the number of classes is 19. The purpose is to classify each activity based on the input features.
 
\subsubsection{Blockchain and Smart Contract}
 
The on-chain smart contracts run on Ganache, a virtual blockchain network \cite{ganache} which runs a local Ethereum blockchain. Furthermore, all smart contracts are written in Solidity language.

\subsubsection{ZKP}

We implement our zkSNARK circuits in Circom2 language with Groth16 as the base algorithm \cite{circom2}.

\subsubsection{Simulation parameters}
To lessen the communication cost, each client combines all weights and biases in all layers into one matrix. Figure 4 demonstrates the combination of parameters of three layers. In addition, to convert floating point to a positive integer for adaptation with the Circom2 language, all amounts are multiplied by a large integer number, and then a large integer offset is added to them.

To analyze the performance of our proposed scheme, we set some parameters at the initialization step. The number of clients is 50, and the number of participating clients in each round varies from 2 to 50. To understand the impact of model complexity, five architectures with different layers and various numbers of neurons in each layer are simulated according to Table 1.

\begin{figure}[!t]
\centering
\includegraphics[width=2in]{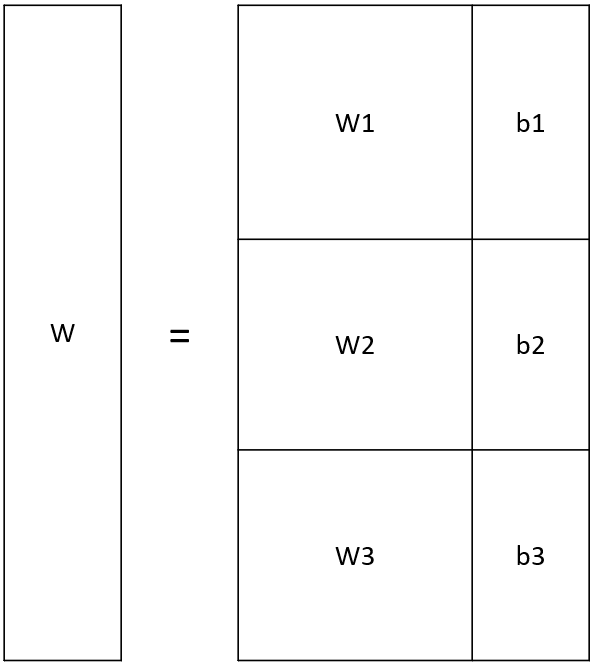}
\caption{Weight combination.}
\label{wights}
\end{figure}

\begin{table}[!t]
\caption{Model Architectures\label{tab:table1}}
\centering
\begin{tabular}{|c|c|c|c|c|c|}
\hline
  & \multicolumn{5}{|c|}{Number Of Neurons} \\
\hline
 Model & Layer1 & Layer2 & Layer3 & Layer4 & Layer5\\
\hline
 Model1& 10 & - & - & - & -\\
\hline
 Model2& 10 & 20 & - & - & -\\
\hline
 Model3& 10 & 20 & 15 & - & -\\
\hline
 Model4& 15 & 20 & 30 & 20 & -\\
\hline
 Model5& 20 & 30 & 40 & 20 & 10\\
\hline
\end{tabular}
\end{table}

\subsection{Model accuracy}
First of all, to illustrate the performance of the model, we conducted experiments with seven client sizes of 2, 5, 10, 20, 30, 40, and 50, and five batch sizes of 10, 25, 50, 75, and 100. Then, we consider 80 percent of the dataset as training data and 20 percent as test data. The training data is divided into 50 segments, and each segment is devoted to each client.

\begin{figure}[!t]
\centering
\includegraphics[width=3in]{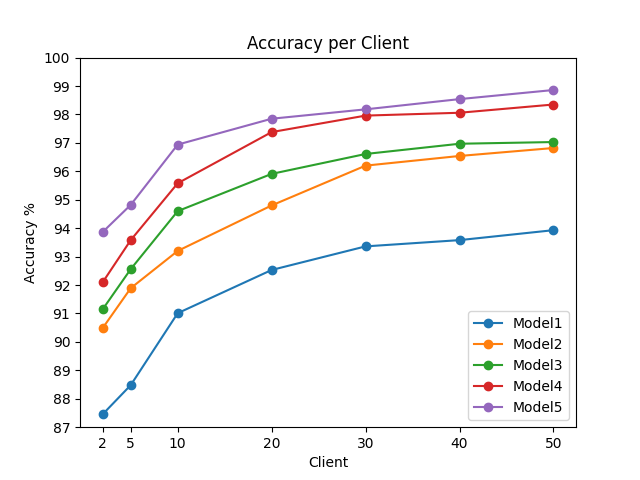}
\caption{Accuracy for different Model Architectures and various number of clients.}
\label{accuracy-node}
\end{figure}

Figure \ref{accuracy-node} illustrates the total accuracy scores achieved for various numbers of clients. In this figure, the accuracy is evaluated for different model architectures. As expected, with the increase in model complexity, the accuracy also increases. On the other hand, as the number of clients grows, the accuracy also rises. The accuracy sees a rapid surge when the number of clients increases from 2 to 20, and after that, the speed of increase in accuracy declines, especially after 30 clients. With 2 clients, the  accuracy is about 87\% for Model1 and around 94\% for Model5. These values rise to around 93\% for Model1 and about 98\% for Model5. The accuracy with 20 nodes is approximately 96\%, 97\%, and 98\% for Model3, Model4, and Mode5 respectively.

\begin{figure}[!t]
\centering
\includegraphics[width=3in]{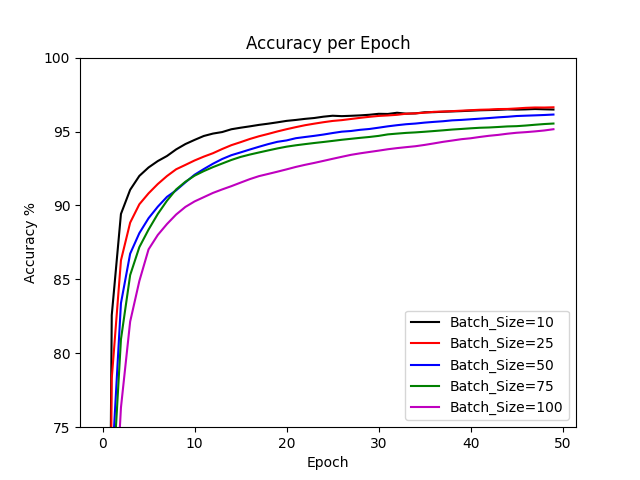}
\caption{Accuracy for different Batch Sizes and various number of epochs.}
\label{accuracy-epoch-batch}
\end{figure}

Figure \ref{accuracy-epoch-batch} shows the total accuracy scores achieved for different batch sizes and various epochs while using the Model2 architecture and 10 clients. According to this figure, accuracy rises in all cases as the number of epochs increases and sees a significant rise up to 20 epochs. After that, the speed of increase lessens. On the other hand, as the batch size decreases from 100 to 10, accuracy rises, with a batch size of 10 achieving the best result compared to other batch sizes, reaching between around 92\% and 95\%  for all batch sizes with 20 epochs. Furthermore, batch size 100 has the lowest accuracy among the five batch sizes.

\begin{figure}[!t]
\centering
\includegraphics[width=3in]{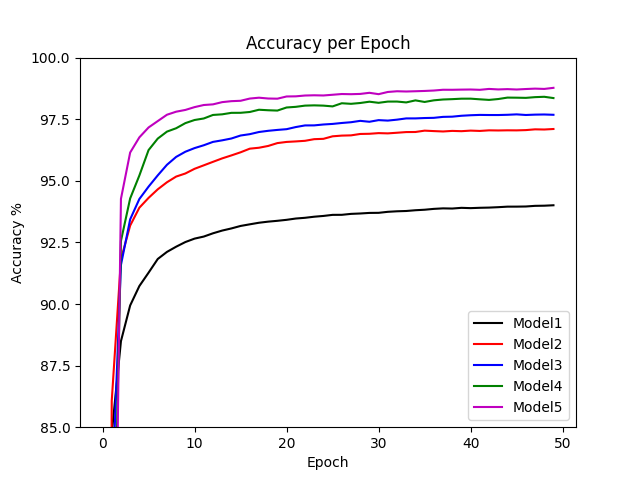}
\caption{Accuracy for different Model Architectures and various number of epochs.}
\label{accuracy-epoch-model}
\end{figure}

The accuracy for different model architectures and various numbers of epochs is seen in Figure \ref{accuracy-epoch-model}. These results are evaluated with 20 clients and a batch size of 10 during 50 epochs. The accuracy sees a sharp rise up to 20 epochs for all models, and after that, it increases slightly. According to this figure, using complex models increases accuracy. However, as we observe, although Model5 has more complexity than Model4, the accuracy doesn't enhance significantly. This observation is important when considering the higher cost of a more complex model for the system, as we will show in the next results.

In the next experiments, we set the batch size to 10, and all figures show the result of one aggregation process in one epoch.

\subsection{ZKP analysis}
\begin{figure}[!t]
\centering
\includegraphics[width=3in]{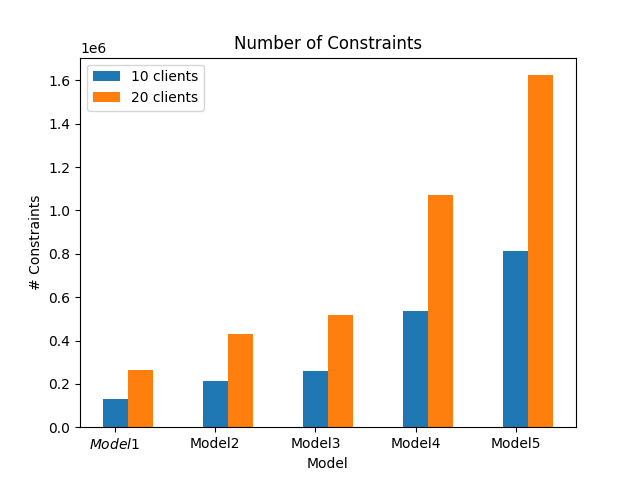}
\caption{Number of Constraints for different Model Architectures.}
\label{constraints}
\end{figure}

One of the most critical factors influencing performance when using ZKP is the proof generation cost. Although the server performs the majority of computations, generating proof associated with the aggregation algorithm incurs a cost. The number of constraints for generating proper proofs for both 10 clients and 20 clients is shown in Figure \ref{constraints}. The number of constraints varies from around 113,000 for Model1 to around 812,000 for Model5 with 10 clients. As expected, the number of constraints experiences a significant increase with the complexity of the model. Furthermore, with the participation of 20 clients, the number of constraints rises to about 226,000 and 1.6 million constraints for Model1 and Model5, respectively. Aggregating more parameters means that more statements must be converted to zkSNARK circuits.

As explained in Section 2, zkSNARK algorithms require substantial computations to convert each statement of the aggregation algorithm to a proper quadratic arithmetic program. By increasing the number of clients, the total number of statements experiences a rise as well.

\begin{figure}[!t]
\centering
\includegraphics[width=3in]{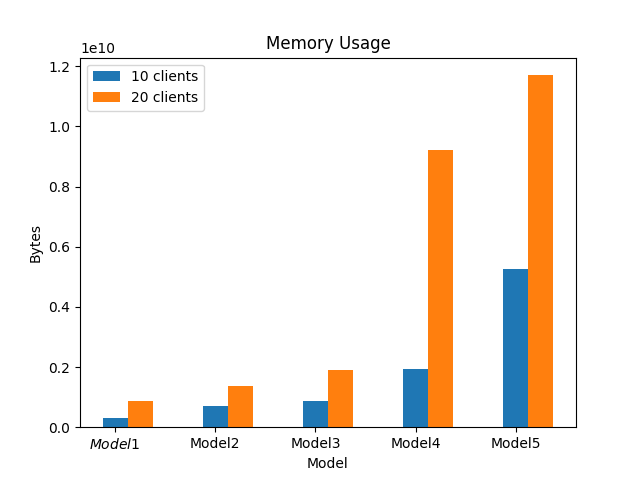}
\caption{Memory usage of different Model Architectures.}
\label{memoryusage}
\end{figure}

The increase in the number of constraints requires more memory to address total computations. This parameter is shown in Figure \ref{memoryusage} for five models with 10 and 20 clients. According to this figure, memory usage rises from about 300 MB for Model1 to about 5.25 GB for Model5 when 10 clients are engaged. On the other hand, when the number of clients is 20, we observe a considerable surge in memory usage. In this scenario, memory consumption rises from 860 MB for Model 1 to around 12 GB for Model5. One noticeable outcome of this figure is that with the use of models with more complexity like Model4 and Model5, there is a rapid rise in memory usage, demonstrating the impact of reaching better accuracy on the amount of costs.

\subsection{Transaction cost analysis}
\begin{figure}[!t]
\centering
\includegraphics[width=3in]{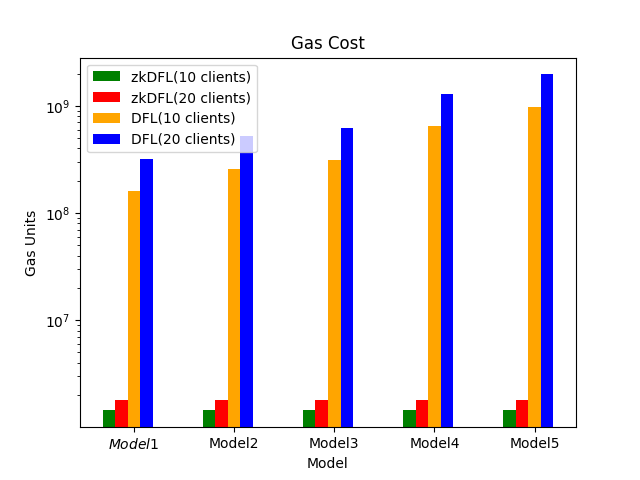}
\caption{Gas cost for different Model Architecture.}
\label{gasCost}
\end{figure}

AAnother influential factor in blockchain-based federated learning is the cost of gas usage, both for deploying and calling smart contracts by the clients or the server. Figure \ref{gasCost} illustrates the gas cost of the proposed work compared to traditional decentralized FL (DFL) systems. In traditional DFL, a smart contract is responsible for aggregating the model parameters, and each client sends its parameters to the smart contract. As shown in Figure \ref{gasCost}, the total gas cost increases sharply from 160 million units to 980 million units with the rise in the number of parameters or model complexity when 10 clients work together. In addition, when the number of clients reaches 20, this cost increases significantly, reaching 320 million units for Model1 and 2 billion units for Model5. In fact, as the number of clients or model complexity grows, the smart contract needs to compute more data on the blockchain, and each extra computation consumes more gas, which is very challenging to execute or even impossible practically.

On the other hand, in our work, the server calculates all computations, and the smart contracts handle only a small proportion of the aggregation task. As seen in Figure \ref{gasCost}, the gas cost is about 1.4 million units when 10 clients are engaged and 1.8 million when 20 clients participate in the training process. Furthermore, a notable result from this figure is that with the increase in model complexity, the gas cost remains steady, since the majority of computations are performed by the server.

By comparing gas costs in zkDFL with traditional DFL systems, we realize that the problem of gas cost for large-scale models in DFL can be addressed with the help of ZKP according to our proposed scheme.

To sum up, while the use of ZKP imposes some additional load on the server, it remarkably reduces the gas cost, which is more crucial for practical federated learning systems.

\section{conclusion}
In this paper, we have introduced a decentralized federated learning framework wherein the aggregation task is performed by a trusted server. The server initiates a zero-knowledge proof (ZKP) protocol to ensure the adherence of the aggregation process to the claimed algorithm. Moreover, the server provides proof that it has utilized all clients' data. Subsequently, clients can verify the validity of the process by calling the deployed smart contracts. Through the evaluation of our implementation, encompassing accuracy, gas cost, and proof generation cost, we conclude that employing a central server with ZKP can effectively address concerns related to trusting the server in centralized FL systems, as well as mitigate the challenges of high gas costs and privacy preservation in decentralized FL systems. This represents a significant step towards practical federated learning systems.

\bibliographystyle{IEEEtran}
\typeout{}
\bibliography{Reference}

\end{document}